\pdfoutput=1
\documentclass[10pt,conference]{IEEEtran}
\IEEEoverridecommandlockouts

\usepackage{cite}
\usepackage{amsmath,amssymb,amsfonts}
\usepackage{graphicx}
\usepackage{textcomp}
\usepackage{xcolor}
\usepackage{url}
\usepackage{microtype}
\usepackage{colortbl}
\usepackage{booktabs}
\usepackage{tabularx}
\usepackage{listings}
\usepackage{multirow}
\usepackage{tikz}
\usepackage[most]{tcolorbox}
\usetikzlibrary{arrows.meta,positioning,calc}
\definecolor{chgcol}{HTML}{C2185B}
\definecolor{clblue}{RGB}{0,84,200}

\newcommand{\para}[1]{\par\smallskip\noindent\textbf{#1}\hspace{0.4em}}

\definecolor{grouprow}{gray}{0.92}
\definecolor{catrow}{gray}{0.80}

\newcolumntype{L}[1]{>{\raggedright\arraybackslash}p{#1}}

\newcommand{\corpussize}{574}
\newcommand{\pilotsize}{148}
\newcommand{\mainsize}{426}

\newcommand{\analysisN}{433}

\newcounter{finding}

\definecolor{findnavy}{HTML}{213C51}
\newtcolorbox{findingbox}[1][]{enhanced, colback=black!5, colframe=findnavy,
  boxrule=0pt, leftrule=2.5pt, arc=0pt, outer arc=0pt,
  left=5pt, right=5pt, top=3pt, bottom=3pt,
  before skip=6pt, after skip=6pt, fontupper=\small, #1}

\newcommand{\obs}[1]{\par\smallskip\noindent\textbf{#1}\enspace}
\definecolor{alert}{HTML}{FF7444}

\definecolor{fwcol}{HTML}{0B6E4F}   %
\definecolor{pfcol}{HTML}{1D63ED}   %

\usepackage{hyperref}
\newcommand{\cobl}[1]{{\color{findnavy}\textbf{#1}}{\,\scriptsize\textcolor{findnavy!70!black}{}}}
\newcommand{\cevent}[1]{{\color{pfcol}#1}{\,\scriptsize\textcolor{pfcol!70!black}{}}}
\newcommand{\creq}[1]{{\color{fwcol}#1}{\,\scriptsize\textcolor{fwcol!70!black}{}}}
\newcommand{\protrowV}[3]{\rowcolor{catrow}\multicolumn{8}{@{}l}{\rule[-0.9ex]{0pt}{3.0ex}\textbf{\MakeUppercase{#1}} (#2 contracts)\;---\;#3}\\}

\begin{document}

\title{Built to Remember, Shipped to Forget: Operational Contracts, Failures, and Repairs in Containerized LLM Agent Deployments}

\author{%
\IEEEauthorblockN{Meriem Mastouri}
\IEEEauthorblockA{\textit{University of North Carolina Wilmington}\\
Wilmington, NC, USA\\
mastourim@uncw.edu}
\and
\IEEEauthorblockN{Emna Ksontini}
\IEEEauthorblockA{\textit{University of North Carolina Wilmington}\\
Wilmington, NC, USA\\
ksontinie@uncw.edu}
\and
\IEEEauthorblockN{Amine Barrak}
\IEEEauthorblockA{\textit{Oakland University}\\
Rochester, MI, USA\\
aminebarrak@oakland.edu}}

\maketitle

\begin{abstract}
LLM agent systems reach production as container images run by platforms
such as Docker, Compose, and Kubernetes. Framework and platform document
opposite expectations: to the platform a container is disposable, while
the framework persists conversations, pauses runs for human input, and
resumes interrupted work from saved state; each side specifies only its own
behavior, so the obligations their composition creates are stated
nowhere. We present the first empirical study of this boundary, stating
these obligations as \emph{operational contracts}: guarantees a
deployment must provide to preserve an agent system's required behavior
when a deployment-platform event occurs. From the first-party
documentation of LangGraph and the three platforms we derive 14
contracts and evaluate them with three independent sources of evidence:
433 deployed open-source agent systems, seven events induced on 12 of
them, and 122 developer repairs mined from 39,261 commits, pull
requests, and issues. At least one contract applies to 75.5\% of the
systems, yet its guarantee is routinely missing: of systems that must
keep state beyond the process,
37.4\%
lose it to a restart and
41.5\%
to a container replacement, and every assessable deployment stops the framework’s production server mid-drain.
Induced events turn 14 of 16 predicted violations into runtime
failures and all
seven predicted holds pass. The failures are born in deployment but
repaired in code: developer repairs independently cover twelve of the
fourteen contracts, 78\% change application code, and only two in three
fully restore the violated guarantee. The 14-contract catalogue,
detection rules, fault-injection harness, and 35-pattern repair
taxonomy provide a basis for checking and strengthening agent
deployments.
\end{abstract}

\section{Introduction}
\label{sec:intro}

Industry surveys report that a majority of organizations are adopting
LLM agents, with over a third already running them in
production~\cite{mckinsey2025stateofai,googlecloud2025roi}.
An \emph{agent system} is an application in which an LLM plans, calls tools, and acts over multiple steps on a user's
behalf, and it reaches its users the way any service does: packaged as
a container image and operated by a platform such as Docker, Compose,
or Kubernetes~\cite{cncfkagent,jebari2026mlbox}.
Agent frameworks endorse this path and instruct developers to package
agents as Docker images~\cite{LSStandalone}. Yet empirical studies of
agent development consistently rank deployment among its
hardest steps~\cite{chen2025llmdev,pan2026measuring}.

One reason is that the platform and the framework document opposite
expectations about the running container. To the platform, the
container is disposable, designed to be stopped and replaced during normal operation, with
nothing of lasting value inside: a stop sends SIGTERM and kills the process
after a grace period of ten seconds by
default~\cite{DockerStop},
an update replaces the container and discards its writable
filesystem~\cite{DockerStorage,ComposeDown}, and anything that must
survive these events is expected to live in volumes outside
it~\cite{TwelveFactor,K8sVolumes}. To the framework, execution is
anything but disposable: agents carry conversations across
requests through a checkpointer~\cite{LGPersistence}, pauses a run
indefinitely until a person approves the next
action~\cite{LGInterrupts}, and resumes interrupted work from saved
state, and warns that resumption may re-execute steps completed before the interruption~\cite{LGDurableLegacy}, potentially repeating external actions such as sending an email, charging a payment, or provisioning a cloud resource. Each side specifies only its
own layer, so whether a deployed agent satisfies both at once is
settled by its artifacts.

Research has approached this junction from two directions. Empirical
studies of deployment artifacts quantify defects, smells, and
misconfigurations in Dockerfiles, Compose files, and
infrastructure-as-code, and build detectors for
them~\cite{cito2017docker,henkel2020binnacle,rosa2024dockerfilesmells,ibrahim2021compose,rahman2019sins,rahman2023k8smisconfig};
the rules they check encode platform best practices that apply to any
containerized service, independent of the application inside.
Empirical studies of LLM and agent applications characterize bugs,
failures, and development challenges within the
application~\cite{chen2025llmdev,pan2026measuring,islam2026agentbugs,cemri2025mast,agentchallenges2025}.
However, no work identifies the obligations that arise only when one layer’s requirements are composed with the other’s lifecycle semantics. Consequently, we do not know what containerized deployments must guarantee for agent systems, how often real deployments provide these guarantees, whether violations become observable failures at runtime, or how developers repair them.

In this paper, we present, to the best of our knowledge, the first
empirical study of the operational boundary between an agent framework
and container platforms. We introduce the \emph{operational contract
(OC)}: a guarantee that a containerized deployment must provide to
preserve an agent system's required behavior when a deployment-platform
event occurs. We select LangGraph for two reasons: it belongs to the most widely adopted open-source toolchain for LLM application development ~\cite{langchain2024,langgraph}, an ecosystem in which 57\% of surveyed teams report running agents in production ~\cite{langchain_state2025} and it explicitly
supports containerized deployment~\cite{LSStandalone}. Our study follows the boundary from specification to practice. We first
derive operational contracts from the first-party documentation of the
framework and deployment platforms, then examine them through three
independent sources of evidence: deployment artifacts from
\analysisN{} open-source agent systems, controlled executions that
induce seven deployment-platform events on 12 systems, and developer
repairs mined from repository histories. We make four contributions:

\begin{itemize}
    \item We derive a catalogue of 14 operational contracts from first-party documentation.

    \item We assess the contracts on 433 deployed open-source agent systems: at least one applies to 75.5\%, and violation rates range from 11\% to 100\%.

    \item We confirm the violations at runtime: under fault injection on 12 systems, 14 of 16 predicted violations manifest as failures, and all seven predicted holds pass.

    \item We study how developers repair violated contracts: from 39,261 mined commits, pull requests, and issues, we identify 122 repairs and extract 35 recurring fix patterns
    
\end{itemize}

The contract catalogue provides practitioners with concrete deployment requirements and tool builders with specifications for automated checks. The repair taxonomy identifies guarantees developers implement in application code, suggesting stronger framework and platform defaults.

\textbf{Replication Package.} All material, prompts and data used in our study are available in our replication package~\cite{MSR2026_replication}.

\begin{figure*}[t]
\centering
\includegraphics[width=0.98\linewidth]{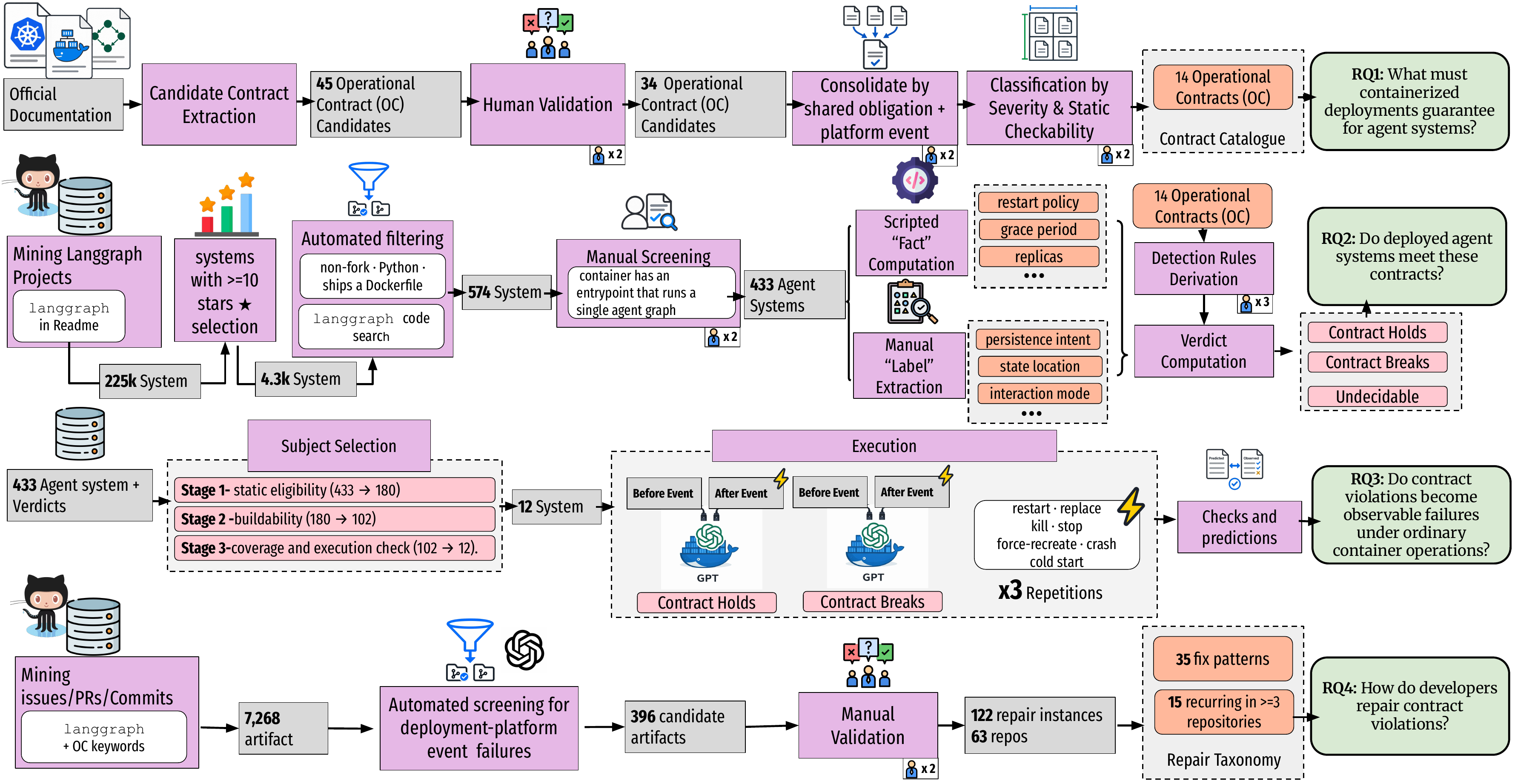}
\caption{Overview of the Study Design}
\label{fig:approach}
\end{figure*}

\section{Study Design}
\label{sec:design}
Figure~\ref{fig:approach} summarizes our study design across the four
research questions. RQ1 draws on documentation and RQ4 on repository
histories; their methods are described in Sections~III and~VI,
respectively. RQ2 and RQ3 study deployed agent systems; this section
describes their shared corpus and the evidence collected from each
system.

\subsection{Corpus extraction}
\label{sec:corpus}
We construct a corpus of LangGraph Agent systems that ship
container deployments. Searching GitHub for repositories whose README
mentions \emph{langgraph} yields 225{,}367 candidates. To focus on
established projects and exclude trivial or personal repositories, we
retain those with at least 10 stars, a threshold used in prior
work~\cite{kalliamvakou2016promises}, leaving
approximately 4{,}345 repositories. From these candidates, we retain non-fork Python repositories that
(i)~use LangGraph in their own code, evidenced by a manifest dependency,
a LangGraph import, or a \texttt{langgraph.json} and (ii)~ship a Dockerfile. These
criteria yield \corpussize{} repositories, which we pin to commits no
later than our 6~July 2026 snapshot cutoff. We then manually verify that
each shipped deployment runs a single agent graph defined in the
repository itself; the following section describes this validation and
the resulting analysis set.

\subsection{Annotation procedure}
\label{sec:instrument}
\label{sec:reliability}

For each system, we collect two forms of evidence. \emph{Facts} are
extracted mechanically from source and deployment files and retain their
file-and-line provenance; examples include whether the system streams
output, its restart policy, and its stop grace period. \emph{Labels}
capture properties that require interpretation of the code. Authors
assign them at the pinned commit using a codebook that defines each
field and its values, records supporting file-and-line evidence, and
distinguishes \emph{absent}, \emph{not applicable}, and
\emph{uncertain}. Some labels apply to the entire corpus; others are
needed only for systems to which a particular contract applies. We call
the latter \emph{targeted readings}. The codebook, and
collected evidence are included in \cite{MSR2026_replication}.

\para{Labeling procedure.}
We establish reliability in two stages. First, we draw a
star-stratified pilot of \pilotsize{} repositories and have two authors
independently assign each applicable label. They adjudicate
disagreements against the code and record the rationale; disagreements
that expose an ambiguous rule lead to a codebook revision and
re-coding of affected systems. Once the codebook is stable, we freeze it
and divide the remaining \mainsize{} repositories between the two
authors. For corpus-wide labels, both authors independently code a
random 10\% of this remaining set to verify that agreement persists.
Targeted readings involve smaller subsets; after establishing pilot
agreement, the first author codes the remaining systems. We
report agreement with the RQ that first uses each label; unless stated
otherwise, reported $\kappa$ values refer to the pilot.

\para{Manual screening: scope and unit of analysis.}
The first label, \emph{scope}, completes corpus selection. While the
automated filters of Section~\ref{sec:corpus} establish that a repository
uses LangGraph and ships a container deployment, scope determines
whether that deployment actually executes a LangGraph graph defined in
the repository. We trace the container entrypoint, Compose services, and
\texttt{langgraph.json} target to the code they execute, excluding
repositories in which no container executes the graph, the graph is not
invoked on the deployed path, or the deployment runs only a model server
or a graph defined elsewhere. A second label establishes the unit of analysis. For each in-scope
repository, we identify its \emph{primary deployed graph} from the
\texttt{langgraph.json} target or, when absent, the graph invoked by the
container entrypoint. Repositories that deploy several unrelated graphs
are labeled \emph{collections} and excluded, avoiding both the
aggregation of unrelated systems into one verdict and multiple systems
from the same repository. Following the labeling procedure above, the authors agree in the pilot
on scope for 137 of 148 repositories ($\kappa=.62$, $AC_1=.91$) and on
collection status for 140 ($\kappa=.61$, $AC_1=.94$). Of the
\corpussize{} repositories, 82 are out of scope and 59 are collections,
leaving \analysisN{} deployed agent systems for analysis.

\section{RQ1: What must containerized deployments guarantee for agent systems?}
\label{sec:rq1}

\para{Motivation.}
Container platforms routinely restart, replace, and terminate
containers, while agent frameworks require state and running work to
survive these deployment-platform events
~\cite{TwelveFactor,burns2016borg,DockerStorage,K8sVolumes,LGPersistence,LGInterrupts,LGDurableLegacy}.
Preserving state alone does not satisfy these requirements: an
interrupted run may remain unfinished even when its state survives, and
resuming it may re-execute side effects
~\cite{LGDurableLegacy}. The framework and deployment platform each
document their own behavior, but neither states the obligation a
deployment must satisfy when the two are combined. RQ1 makes these
obligations explicit as operational contracts.

\begin{table*}[!tp]
\caption{The fourteen operational contracts and their RQ2 verdicts across
the \analysisN{} deployed agent systems. Contracts are grouped by what
they protect. Static indicates whether a contract is checkable from
repository artifacts.}
\label{tab:contracts}
\centering
\scriptsize
\setlength{\tabcolsep}{3pt}
\renewcommand{\arraystretch}{1.12}
\begin{tabular}{@{}l L{8.0cm} L{3.7cm} c c r r r@{}}
\toprule
& & & & & \multicolumn{3}{c@{}}{\textbf{Verdicts (RQ2)}} \\
\cmidrule(l){6-8}
& \textbf{Contract} & \textbf{Applies to} & \textbf{Severity} & \textbf{Static} & \textbf{Apply} & \textbf{Hold} & \textbf{Broken} \\
\midrule
\protrowV{Saved state}{4}{what the agent keeps between requests: conversations, paused runs and memories}
OC1 & The deployment must ensure that \cobl{state outlives the process}, so that \cevent{a restart (crash, OOM kill, failed probe)} cannot violate \creq{the availability of saved threads, paused runs and memories}. & Systems that keep threads, paused runs or memories across requests & High & $\checkmark$ & 265 & 166 & \textbf{99 (37.4\%)} \\
OC2 & The deployment must ensure that \cobl{state outlives the container}, so that \cevent{a replacement (redeploy, \texttt{down}/\texttt{up}, update)} cannot violate \creq{the availability of state saved for later requests}. & Systems that keep state across requests and are redeployed or updated & High & $\checkmark$ & 265 & {155} & \textbf{{110 (41.5\%)}} \\
OC3 & The deployment must ensure that \cobl{every replica reaches the same state}, so that \cevent{routing across several replicas} cannot violate \creq{one view of each thread}. & Systems that keep state across requests and run several replicas & High & $\checkmark$ & {19} & {16} & \textbf{{3 (15.8\%)}} \\
OC10 & The deployment must ensure that \cobl{the database accepts connections before setup}, so that \cevent{start ordering} cannot violate \creq{the start-up migrations}. & Systems that start together with their database & Medium & $\checkmark$ & 46 & {36} & \textbf{{6 (14.3\%)}} \\
\protrowV{Running work}{8}{what the agent is doing: a run in progress and the workers that execute it}
OC4 & The deployment must ensure that \cobl{the grace period covers the drain}, so that \cevent{a stop (SIGTERM, then SIGKILL)} cannot violate \creq{the drain of in-flight runs}. & Systems that drain runs on shutdown & Medium & ($\checkmark$) & {27} & 0 & 8 (100\%) \\
OC5 & The deployment must ensure that \cobl{SIGTERM reaches the application}, so that \cevent{a stop (SIGTERM sent only to the container's PID 1)} cannot violate \creq{the drain that starts on SIGTERM}. & Systems that drain on SIGTERM from their own image & Medium & $\checkmark$ & {15} & {12} & \textbf{{3 (20.0\%)}} \\
OC6 & The deployment must ensure that \cobl{progress is saved during the run}, so that \cevent{a kill, OOM kill or hard eviction} cannot violate \creq{the steps completed so far}. & Systems with long multi-step runs and a durable checkpointer & Medium & $\checkmark$ & 9 & 8 & 1 (11.1\%) \\
OC7 & The deployment must ensure that \cobl{re-executed side effects are harmless}, so that \cevent{a kill in the middle of a step} cannot violate \creq{actions that must happen once}. & Systems that act on the world and resume or retry & High &  & 87 & --- & --- \\
OC8$^\ddagger$ & The deployment must ensure that \cobl{interrupted runs are resumed}, so that \cevent{a kill, restart or replacement} cannot violate \creq{the completion of saved runs}. & Systems that run agents in background tasks & Medium &  & 61 & --- & --- \\
OC12$^\dagger$ & The deployment must ensure that \cobl{the server restarts after a crash}, so that \cevent{a crash under Docker's default \texttt{restart:\,no}} cannot violate \creq{a worker listening at all times}. & Systems that self-host the server on Docker & Medium & $\checkmark$ & 11 & 1 & \textbf{6 (85.7\%)} \\
OC13$^\dagger$ & The deployment must ensure that \cobl{an update keeps a worker running}, so that \cevent{a Recreate or stop-first update} cannot violate \creq{a worker listening at all times}. & Systems that update the server & Medium & $\checkmark$ & 12 & 1 & \textbf{7 (87.5\%)} \\
OC14 & The deployment must ensure that \cobl{the memory limit fits all concurrent runs}, so that \cevent{an OOM kill at the limit} cannot violate \creq{the completion of concurrent runs}. & Systems that set memory limits & Medium &  & 53 & --- & --- \\
\protrowV{Clients and external resources}{2}{what lies outside the run: the user waiting for its output and the resources it pays for}
OC9 & The deployment must ensure that \cobl{releasing external resources does not depend only on shutdown code}, so that \cevent{a kill (no shutdown code runs)} cannot violate \creq{the release of sandboxes and billable resources}. & Systems that create sandboxes or billable resources  & Medium & $\checkmark$ & {22} & 8 & \textbf{{14 (63.6\%)}} \\
OC11$^\ddagger$ & The deployment must ensure that \cobl{streaming clients can resume after a redeploy}, so that \cevent{a replacement with open connections} cannot violate \creq{delivery of the full answer}. & Systems that stream output to a client & Low &  & 155 & --- & --- \\
\bottomrule
\end{tabular}
\vspace{2pt}
\begin{minipage}{\textwidth}
\scriptsize
$\checkmark$: statically checkable;
($\checkmark$): statically checkable only for production-server deployments;
---: requires runtime evidence.
$^\dagger$: production server only;
$^\ddagger$: without production server.
\end{minipage}
\end{table*}

\para{Approach.}
We derive operational contracts by reading the agent-framework and
deployment-platform documentation together. Each contract connects three
elements: an \creq{agent-framework requirement}, a
\cevent{deployment-platform event} that can violate it, and the
\cobl{deployment obligation} that prevents the violation. A contract
applies only to systems for which its agent-framework requirement is
relevant. We derive the
contracts in four steps: source selection, candidate extraction, human
validation and consolidation, and classification.

\underline{(i)~Source selection.}
Following prior work on deriving developer directives from official
documentation~\cite{monperrus2012directives}, we use only first-party
sources. For the agent framework, we include the LangGraph libraries and
production server; for the deployment platform, we include Docker,
Docker Compose, and Kubernetes, covering container execution,
composition, and orchestration
~\cite{DockerStorage,ComposeDown,K8sVolumes,K8sVirtualIPs,K8sJobs,K8sEviction}.
For reproducibility, we freeze each documentation set at the versions used in the study; exact versions and snapshots are in the replication package.

\begin{figure}[t]
    \centering
    \includegraphics[width=0.9\columnwidth]{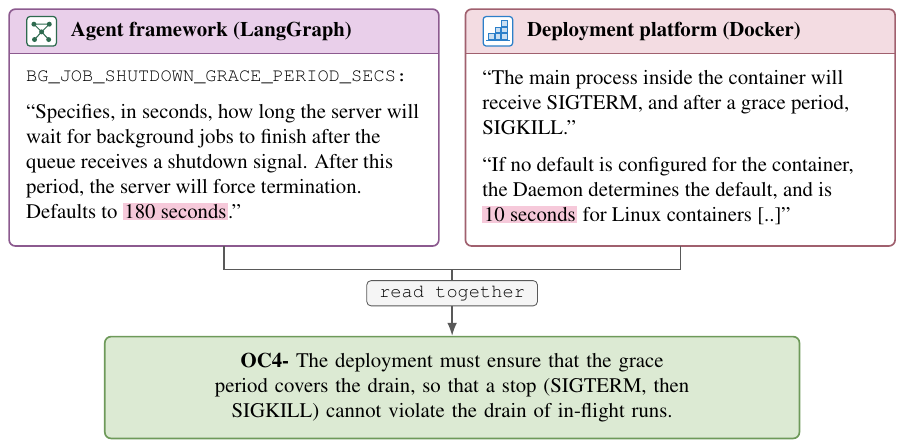}
    \caption{Agent-framework and deployment-platform evidence for a
    candidate operational contract.}
    \label{fig:oc-derivation}
\end{figure}

\underline{(ii)~Candidate extraction.}
A candidate contract connects a documented \creq{agent-framework
requirement} with a documented \cevent{deployment-platform event} that
can violate it. We use Claude Opus~5.5 as an extraction instrument. The
model extracts both statements verbatim with their sources, pairs them
across the two documentation sets, and turns each conflicting pair into
a candidate contract by stating the corresponding \cobl{deployment
obligation}. Figure~\ref{fig:oc-derivation} illustrates the process. LangGraph's
production server allows up to 180 seconds to drain running work after a
shutdown signal~\cite{LSEnvVars}, whereas Docker sends SIGTERM on stop
and SIGKILL after a default 10-second grace period~\cite{DockerStop}.
Together, these statements yield the candidate contract
\cobl{the stop grace period covers the drain}, so that \cevent{a stop}
does not violate \creq{the drain of running work}. A script verifies
every extracted quotation against the pinned documentation, and two
authors validate every candidate in step~(iii). This use of an LLM for
documentation extraction, followed by human validation, follows prior
work~\cite{meng2024chatafl,xie2025llmspecs,li2025formalspecs,ahmed2025llmannotation}.
The prompt and raw model output are available in
the replication package~\cite{MSR2026_replication}.

\underline{(iii)~Validation and consolidation.}
Two authors independently validate the 45 candidates against their
supporting agent-framework and deployment-platform quotations. A
candidate is retained only when the documentation establishes an
agent-framework requirement and a deployment-platform
event that can violate it, and the conflict can arise under normal
deployment-platform operation without requiring an exceptional event or
a specialized configuration. The authors agree on 41 of 45 decisions
($\kappa=.72$, $AC_1=.87$); disagreements are resolved against the
quoted evidence. This validation excludes 11 candidates and retains 34. The authors then independently consolidate candidates that express the
same deployment obligation under the same
deployment-platform event. For example, three candidates
require state to survive a container restart but concern different forms
of state: conversation checkpoints in an in-memory checkpointer,
long-term memories in an in-memory store, and LangGraph's
state directory. Because the deployment
obligation and deployment-platform event are the same, we consolidate
them into one operational contract and retain the three forms of state
as instances; the contract applies only to systems that maintain the
corresponding state. The authors agree on 94.7\% of pairwise
consolidation decisions ($\kappa=.71$); after adjudication,
consolidation yields 14 operational contracts.

\underline{(iv)~Classification.}
Finally, we classify each operational contract by \emph{severity} and
\emph{static checkability}. Severity captures the recoverability of the
documented consequence~\cite{Elnozahy2002,Helland2007}: \emph{high} if
retry, restart, or manual intervention cannot restore the intended
outcome, as when saved state is lost or an external action is repeated;
\emph{medium} if execution is disrupted but recovery remains possible,
as when running work is lost or stalled, a service remains unavailable,
or resources remain running until corrective action; and \emph{low} if
the effect is recoverable and confined to a single request or stream. Static checkability captures whether a contract can be evaluated from
repository artifacts alone. A contract is \emph{statically checkable}
when both its applicability and the evidence needed to determine whether
its deployment obligation is satisfied are observable in source and
deployment files; otherwise, runtime evidence is required. The authors
fixed these criteria before classification and independently classified
all 14 contracts, agreeing on all 14 severity classifications and 13 of
14 static-checkability classifications. The disagreement was
resolved by discussion. 

\para{Results.} Table~\ref{tab:contracts} lists the 14.

\obs{Deploying agents on container platforms gives rise to fourteen operational contracts, each guarding against a distinct failure risk.}
Saved state can be lost when a process restarts (OC1), when its container
is replaced (OC2), or become inconsistent when requests reach replicas
that do not share state (OC3). Running work introduces different
conflicts: a stop can terminate a run before its drain completes (OC4),
while resumption after a mid-step interruption can repeat an external
action (OC7). The consequences extend beyond the run: external resources
can remain running after a kill (OC9), and container replacement can
interrupt a client's stream (OC11). In each case, the framework
and platform behave as documented; the violation arises from
their composition.

\obs{Persisting agent state satisfies only four of the fourteen deployment
obligations.}
The other ten extend beyond saved state: running work must drain or
resume after interruption, re-executed side effects must be harmless,
workers must remain available, interrupted streams must resume, and
external resources must be released.

\obs{Four contracts protect against failures whose effects no retry or
restart can undo.} The high-severity contracts are OC1--OC3, where saved state is lost,
and OC7, where re-execution repeats an external action. 

\obs{More than half of the contracts are statically checkable before agent deployment.} These include all four saved-state contracts (OC1--OC3, OC10) and three
of the four high-severity contracts. The remaining contracts require
runtime evidence that repository artifacts do not expose: OC7 depends
on whether re-executed side effects are harmless, while OC4 is
statically checkable only for LangGraph's production server, whose drain
window is explicitly configured.

\section{RQ2: Do Deployed Agent Systems Meet the Contracts?}
\label{sec:rq2}

\para{Motivation.}
RQ1 identified fourteen operational contracts for containerized agent
deployments, nine of which are statically checkable. Whether they matter
in practice depends on their applicability and on whether deployments
satisfy their obligations. RQ2 therefore examine the \analysisN{}
systems in our corpus to determine which contracts apply
and which are violated.

\para{Approach.}
RQ2 uses the two forms of evidence defined in
Section~\ref{sec:instrument}: \emph{facts}, extracted from source and
deployment files, and \emph{labels}, assigned by the authors under the
procedure of Section~\ref{sec:reliability}. We use facts when repository
evidence is explicit and labels when interpretation is required. Each labeled
field is defined at its first use, with its agreement. For each system, we determine whether a contract \emph{applies} and,
when it does, whether it \emph{holds}. A contract applies when its
agent-framework requirement is relevant to the system and holds when the
deployment satisfies its obligation. We encode these decisions in one
\emph{detection rule} per contract, derived from
Table~\ref{tab:contracts} and its supporting quotations. One author
derives each rule, a second independently reviews it against the
contract, and a third adjudicates disagreements. The complete rules are in \cite{MSR2026_replication}.

\underline{(i)~Contract Applicability evidence.} Applicability is mostly
decided by  \emph{facts}: streaming, background workers, interrupts, replicas,
and the use of LangGraph's production server. Two aspects require
\emph{labels}. \emph{Interaction mode} ($\kappa=.71$) characterizes how the
system meets its workload: a multi-turn conversation, a long-running
job, or a stateless request. \emph{Persistence intent} ($\kappa=.71$)
captures whether the system is designed to carry state across
requests, evidenced by threads, checkpointers,
a human-in-the-loop, or explicit reloading of prior history;
such system \emph{needs durable state}, and the state contracts
apply to it.

\underline{(ii)~Contract satisfaction evidence.}
For some contracts, \emph{facts} directly establish whether the
deployment obligation is satisfied: the restart policy (OC12), stop
grace period (OC4), update strategy (OC13), volume for database data
(OC2), run durability setting (OC6), and wait for database health
(OC10). The remaining require \emph{labels}. For the saved-state contracts (OC1--OC3), we first label the
checkpointer used by the deployed system ($\kappa=.73$), then inspect
cases where persistence depends on application or deployment behavior:
where 83 systems without a checkpointer keep their state
($\kappa=.78$); which checkpointer the deployment selects when several
can be constructed (63 systems; $\kappa=.72$); whether
checkpoint files lie within a declared mount
(63; $\kappa=.73$); and whether paused runs
survive restart (77 systems; $\kappa=.94$). For 135 systems that start
several containers from one image, we additionally determine whether
they are replicas or distinct roles ($\kappa=.83$) and whether every
instance reaches the same state ($\kappa=.72$). For OC8 and OC9, we inspect whether the repository provides a mechanism
to resume interrupted runs and to release external resources after a
kill (93 systems; $\kappa=.85$--$1.00$ on the 28 double-coded systems).
For (OC2--OC6),
we additionally search the pinned source for the
implementations relevant to each obligation, including persistent
volumes for Kubernetes-hosted state, worker processes, shutdown
handlers, and progress saved during a run, and inspect every match
($\kappa=.78$).

\underline{(iii)~Verdicts.}
A script applies the detection rules to the collected facts and labels,
assigning each system one verdict per contract: \emph{does not apply},
\emph{holds}, \emph{broken}, or \emph{not decidable}. A contract is
\emph{broken} only when repository evidence establishes that the
deployment does not satisfy its obligation; it is \emph{not decidable}
when repository evidence is insufficient. OC7, OC8, OC11, and OC14
require runtime evidence and therefore receive no repository verdict.
OC4 is decidable only for systems using LangGraph's production server.

\para{Results.} Table~\ref{tab:contracts} presents the verdict for every contract.

\obs{At least one operational contract applies to 75.5\% of agent
systems: 61.2\% need durable state, 19.9\% run work in the background,
and 60.3\% stream output to clients.}
More than half of the systems maintain multi-turn conversations
(54.5\%), and 13.4\% pause runs for human input; together with systems
that maintain memories, 61.2\% need durable state, making the saved-state
contracts applicable (Table~\ref{tab:behav}). OC11 applies to the
155
systems that stream from their own server rather than LangGraph's
production server. Notably, 6.0\% use a single constant thread across
requests, causing all users to share one conversation. Overall, 327 of
the \analysisN{} systems (75.5\%) have at least one applicable
operational contract.

\begin{table}[t]
\caption{RQ2 applicability evidence across the \analysisN{} deployed
agent systems.}
\label{tab:behav}
\centering
\footnotesize
\setlength{\tabcolsep}{4pt}
\renewcommand{\arraystretch}{1.08}
\begin{tabular}{@{}lrr@{}}
\toprule
\textbf{Evidence} & $\boldsymbol{n}$ & \textbf{\%} \\
\midrule
\rowcolor{grouprow}
\multicolumn{3}{@{}l}{\textbf{Interaction mode} \hfill \emph{(labels)}} \\
Multi-turn conversation & 236 & 54.5 \\
Stateless request & 113 & 26.1 \\
Batch job & 35 & 8.1 \\
Long-running job & 35 & 8.1 \\
Event-driven & 14 & 3.2 \\
\rowcolor{grouprow}
\multicolumn{3}{@{}l}{\textbf{Workload} \hfill \emph{(facts)}} \\
Streams output to a client & 261 & 60.3 \\
Consumes a queue / runs workers & 86 & 19.9 \\
Pauses for human input & 58 & 13.4 \\
\rowcolor{grouprow}
\multicolumn{3}{@{}l}{\textbf{Persistence intent} \hfill \emph{(labels)}} \\
Durable: carries state, survives a restart & 167 & 38.6 \\
Not durable: carries state, loses it & 98 & 22.6 \\
Needs durable state (durable $+$ not durable) & 265 & 61.2 \\
Stateless by design & 140 & 32.3 \\
In memory, within one request & 28 & 6.5 \\
\bottomrule
\end{tabular}
\end{table}

\obs{37\% of agent systems that need durable state lose it on an
ordinary restart.}
Of the 265 systems to which OC1 applies, 166 (62.6\%) hold the contract,
whereas 99 (37.4\% [31.8--43.3]) lose a conversation, paused run, or
memory that must outlive the process. The rate is similar among
multi-turn conversational systems (87/221, 39.4\%). Most systems that
break OC1 (77/99, 77.8\%) ship no persistent store; the remaining 22
ship one that the agent state does not reach. Indeed, 16 of 76 systems
using an in-memory checkpointer deploy a persistent store alongside it
without routing agent state there.

\begin{figure}[h!]
\centering
\includegraphics[width=.92\columnwidth]{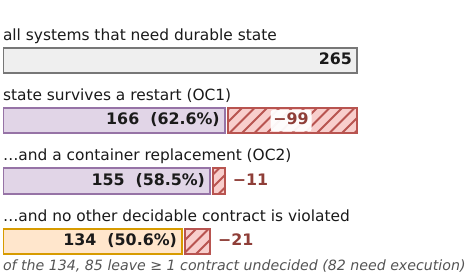}
\caption{265 systems that need durable state across three cumulative checks. Solid bars satisfy all preceding contracts; hatched segments violate the contract at that step.%
}
\label{fig:ladder}
\end{figure}

\obs{Preserving agent state through one deployment-platform event does not
ensure it survives the next.}
Figure~\ref{fig:ladder} follows the 265 systems that need durable state
through cumulative contract checks. Of these, 166 (62.6\%) preserve
state across restart (OC1), and 155 (58.5\%) also preserve it across
container replacement (OC2); thus, eleven survive restart but lose state
on redeployment. Only 134 (50.6\%) have no further broken contract that
repository evidence can establish. This is an upper bound: 85 of the
134 have at least one applicable contract that is not decidable from
repository evidence, including 82 that require runtime evidence.

\obs{Durable agent state does not imply resumable running work:
88.5\% of systems running work in the background ship no mechanism to resume an interrupted run.}
OC8 requires runtime evidence, so we report what the repositories
provide. Of the
61 applicable systems, 54 (88.5\% [78.2--94.3]) provide no
resumption mechanism; six use a queue that redelivers unacknowledged
work, and one resumes unfinished runs at start-up. Notably, 28 of the 54
hold OC1: their saved state outlives the process, but the repository
provides no mechanism to resume the interrupted run.

\obs{Two in five agent systems that pause for human input lose
the paused run on restart.}
Of the 58 systems that pause for human input, two are not decidable from
repository evidence. Among the remaining 56, 23 (41.1\%
[29.2--54.1]) break OC1: the paused run does not outlive the process.
The rate is identical across the two coding blocks ($p=1.0$). Thus, a
restart can discard a run while it is waiting for human
input.

\obs{Every production-server deployment provides less time
than required to drain running work.}
OC4 applies to 27 systems: 12 use LangGraph's production server, with
its 180-second drain window, and 15 implement their own drain. Because
the latter depends on the running work being drained, OC4 is decidable
from repository evidence only for eight production-server deployments.
All eight retain Docker's default 10-second stop grace period, 170
seconds shorter than the configured drain window. Three systems with
their own drain fail earlier: SIGTERM does not reach the application, so
the drain never starts (OC5). Related worker obligations are also
frequently broken: six of seven decidable production-server deployments
break OC12, and seven of eight break OC13.

\obs{A kill leaves an agent's external resources running: 14 of 22
systems that create sandboxes or billable resources release them only
through shutdown code a kill never runs or not at all (OC9).}
Mid-run
progress, by contrast, is rarely at risk: of the {9} batch,
event-driven systems with a durable checkpointer,
{8} keep LangGraph's default per-step saving, which loses at most
the last step on a kill, and {one} disables it
(\texttt{durability="exit"}) by a documented choice (OC6). Where long
runs do lose progress, the cause is state kept in process memory (OC1)
or a run nothing resumes (OC8).

\providecommand{\rqtodo}[1]{\textcolor{red}{\textbf{[TODO: #1]}}}

\section{RQ3: Do the Violations Become Observable Failures Under Ordinary Container Operations?}
\label{sec:rq3exec}

\para{Motivation.}
RQ2 derives verdicts from repository evidence, but cannot show what
happens when a deployment-platform event occurs; some contracts also
require runtime evidence. RQ3 therefore executes the deployments,
induces the deployment-platform events, and observes
whether contracts predicted to hold or be broken behave accordingly at
runtime.

\para{Approach.}
We evaluate the RQ2 verdicts in four steps: (i)~select contracts whose
deployment-platform events can be reproduced and observed, (ii)~select
systems predicted to hold or break each contract, (iii)~deploy each
system as shipped and induce the corresponding deployment-platform
event, and (iv)~compare the outcome with the prediction.

\underline{(i)~Contract and event selection.}
We select contracts whose deployment-platform event can be induced by a
documented Docker or Compose operation and whose outcome can be observed
without modifying the system's code or deployment configuration, apart
from the model-routing changes of step~(iii). The resulting events
(Table~\ref{tab:rq4events}) are restart, container replacement, kill, stop under Docker's
default 10-second grace period, forced recreation, server-process crash,
and cold start.
\footnote{We reproduce a crash by sending SIGKILL directly to the server process,
analogous to an out-of-memory kill. We avoid \texttt{docker kill}
because Docker treats it as a user stop and bypasses the restart policy.}
The expected runtime outcome follows from the documented
deployment-platform semantics. For example, restart clears process
memory but preserves the container filesystem; replacement discards that
filesystem; and named volumes survive both. Before executing any agent
system, we verify each event effect on a minimal
Compose deployment. Eleven of the fourteen contracts meet these criteria. We exclude OC3
because testing it requires adding replicas, OC14 because inducing an
OOM kill requires changing the memory limit or constructing a workload
that exhausts it, and OC9 because its external resources require paid
provider accounts. OC7 and OC11 require runtime evidence, so we execute
them without an RQ2 prediction and report the observed outcome. OC8 has no RQ2
verdict: repository files can reveal a resumption mechanism,
but not whether interrupted runs actually resume. We instead
predict its outcome from the OC8 reading (Section~\ref{sec:rq2}):
a system with no mechanism should leave the run
unfinished, and one that resumes runs at start-up or takes them
from a redelivering queue should complete it.

\underline{(ii)~System selection.}
Executing the full corpus is impractical because each system must be
built, configured, and exercised before a deployment-platform event can
be induced. We therefore select for contract coverage: for each executed
contract with an RQ2 verdict, we seek a system in which the contract is
\emph{broken} and, where available, one in which it \emph{holds}.
Because a system can exercise several contracts, we seek a small set
that covers these cases. Selection proceeds in stages.

\emph{Stage 1: static eligibility (\analysisN{} $\rightarrow$ 180).}
Using the facts of Section~\ref{sec:instrument}, we retain systems that
can be exercised without modifying their code or deployment files. Each
must ship a Compose deployment, so the deployment-platform events can be
induced on the deployment as shipped; expose an HTTP API or LangGraph's
production server, so requests can be issued programmatically; use an
OpenAI-compatible model client, so all systems can be
routed to the same model through configuration; require no GPU; and
have an RQ2 verdict, or for OC8 a reading-based prediction, for at least one executed contract. OC7 and OC11
require no additional systems because we measure them on selected
systems to which they apply.

\emph{Stage 2: buildability (180 $\rightarrow$ 102).}
We retain systems whose images can be built or pulled and require no local model.

\emph{Stage 3: coverage and execution check (102 $\rightarrow$ 12).}
We use greedy set cover to select systems that collectively cover the
required RQ2 verdicts while allowing one system to exercise multiple
contracts. We then run each selected deployment and confirm that it
accepts requests, obtains its model endpoint from the environment, and
requires no paid service; otherwise, we replace it with the next
candidate. This process yields 12 systems
(Table~\ref{tab:rq4results}). When only a \emph{holds} or \emph{broken} case is runnable, we exercise
that case alone. Each contract
is exercised on only one to five systems.

\underline{(iii)~Execution.}
We reproduce each selected system from its pinned commit using the
deployment configuration provided by its developers. We retain the
shipped configuration and change only the model endpoint and credentials required for the
experiment. We record the digest of every container image executed.

All systems use the same pinned model
(\texttt{gpt-4.1-mini-2025-04-14}, temperature~0), accessed through a
recording proxy. The proxy serves two purposes: it removes the model as
a source of variation across systems and records every model request for
the outcome checks in step~(iv).

For contracts that require interrupting a run in progress, we must
ensure that the run has not completed before the deployment-platform
event occurs. The proxy therefore delays the current model request for
at most 60\,s, giving us time to induce the event, and releases the
request afterward. Thus, when the experiment tests an interruption, the
deployment-platform event occurs while the agent run is still in
progress.

\underline{(iv)~Checks and predictions.}
For each RQ2 verdict, and
for OC8 from its reading, we fix the expected runtime outcome before
execution: a contract that is \emph{broken} should fail its check,
whereas one that \emph{holds} should pass. An observation confirms or
contradicts this prediction, or is inconclusive
(Table~\ref{tab:rq4results}). OC7 and OC11 have no prediction and are
measured directly. Each contract is exercised under a workload that makes its obligation
observable. Saved-state contracts use a conversation, running-work
contracts use a run in progress, and OC5, OC10, and OC12 require no
workload because they test an idle stop, cold start, and server crash,
respectively. For saved state, we use LongMemEval~\cite{wu2025longmemeval} to determine
whether a conversation survives without judging the model's answer. We
replay its three shortest single-session histories as separate threads;
each provides a later question and identifies the earlier turns needed
to answer it. We ask the question and inspect the proxy log for those
turns: their presence means the conversation is preserved; their
absence, or a thread-not-found response, means it is lost; a connection
failure is inconclusive. We perform this check after replay, after
waiting for the duration of the deployment-platform event without
inducing it, and after the event. We score post-event loss only when pre-event checks confirm that
the conversation was preserved. For running work, we start a run through the system's API using a
repository-provided request from its README, tests, or example code; the
exact requests are in \cite{MSR2026_replication}. The proxy keeps the run
in progress while the deployment-platform event is induced.
Table~\ref{tab:rq4events} gives the contract-specific checks: running
work drains before termination (OC4), progress is saved during the run
(OC6), and an interrupted run resumes without resubmission within
300\,s (OC8). OC5 checks whether SIGTERM reaches the application, OC10
whether the system becomes ready within 120\,s of a cold start, and
OC12 whether the server restarts within 60\,s of a crash. OC7 and OC11
measure whether re-execution repeats an external action and whether an
interrupted stream can resume, respectively. We repeat each deployment-platform event three times. A repetition
counts only when its required pre-event condition holds: the
conversation is available, the run is in progress, or the server is
running, as appropriate. We report an outcome when at least two
repetitions are valid, and all valid repetitions agree; otherwise, the
outcome is inconclusive.

\begin{table*}[!t]
\caption{RQ3 deployment-platform events, runtime checks, and outcomes
across 12 agent systems.}
\label{tab:rq4events}\label{tab:rq4results}
\centering
\scriptsize
\setlength{\tabcolsep}{4pt}
\renewcommand{\arraystretch}{1.08}

\begin{tabular*}{\textwidth}{
@{\extracolsep{\fill}}
l
L{6cm}
L{6cm}
c
c
@{}}
\toprule
\textbf{Contract} &
\textbf{Deployment-platform event} &
\textbf{Contract holds if} &
\textbf{Broken} &
\textbf{Holds} \\
\midrule

\rowcolor{gray!15}
\multicolumn{5}{@{}l}{\textbf{SAVED STATE}} \\

OC1 (conversation) &
\texttt{restart} &
earlier turns still reach the model &
1/1 & 2/2 \\

OC1 (paused run) &
\texttt{restart} while awaiting input &
the paused run can be resumed &
1/1 & 1/1 \\

OC2 &
\texttt{down}; \texttt{up -d} &
earlier turns still reach the model &
3/4$^{a}$ & -- \\

\rowcolor{gray!15}
\multicolumn{5}{@{}l}{\textbf{RUNNING WORK}} \\

OC4 &
\texttt{stop} while a run is in progress &
running work drains before the container exits &
2/2$^{b}$ & -- \\

OC5 &
\texttt{stop} of an idle service &
SIGTERM reaches the application &
1/1 & 0/1$^{c}$ \\

OC6 &
\texttt{kill} after third model call; \texttt{up -d} &
a finished step is in the checkpoint store &
-- & 1/1 \\

OC8 &
\texttt{kill} while a run is in progress; \texttt{up -d} &
the run completes without resubmission within 300\,s &
2/2 & 1/1 \\

OC10 &
cold first \texttt{up} &
the system is ready within 120\,s &
0/1$^{d}$ & 2/2 \\

OC12 &
SIGKILL to the server process &
the server restarts within 60\,s &
2/2 & -- \\

OC13 &
\texttt{up --force-recreate} while a run is in progress &
the run survives the update &
2/2$^{b}$ & -- \\

\rowcolor{gray!15}
\multicolumn{5}{@{}l}{\textbf{CLIENTS AND EXTERNAL RESOURCES}} \\

OC7 &
\texttt{kill} after an external action; re-send &
the external action is not repeated &
\multicolumn{2}{c}{repeated (1)} \\

OC11 &
\texttt{up --force-recreate} during streaming &
the interrupted stream can resume &
\multicolumn{2}{c}{not resumed (1)} \\

\midrule
\textbf{Total} & & &
\textbf{14/16} &
\textbf{7/7}$^{c}$ \\
\bottomrule
\end{tabular*}

\vspace{2pt}
\begin{minipage}{\textwidth}
\scriptsize
\emph{Notes:} Broken/Holds: systems whose runtime outcome matches the
RQ2 verdict; OC7 and OC11 are measured directly.
$^{a}$~The pulled image preserves conversation state on a named volume.
$^{b}$~The interrupted run later resumes.
$^{c}$~Inconclusive: one valid repetition.
$^{d}$~The image retries the database before serving.
\end{minipage}
\vspace{-0.3cm}
\end{table*}

\para{Results.}  Table~\ref{tab:rq4results} lists every outcome.

\obs{The code predicts what happens to deployed agents: 87.5\% of
predicted violations manifest as runtime failures, while every scorable
prediction that a contract holds is confirmed.}
Fourteen of the 16 predicted violations (87.5\%) manifest as the
corresponding runtime failure. Of the eight systems predicted to
satisfy their contract, seven pass the runtime check; the remaining
case is inconclusive (Table~\ref{tab:rq4results}, note~$c$).

\obs{Execution shows that satisfying one contract does not ensure the
next: agent state can survive a restart but not a replacement, an
interrupted run can retain its state but never resume, and shutdown can
begin but not finish before termination.} \texttt{tool-kit-ai} and \texttt{SRE-Agent} keep all tested conversations
across restart (OC1) but lose them after container replacement (OC2),
because their state remains inside the container. \texttt{langmanus}
keeps conversation state in process memory and loses it on restart
(OC1). The distinction extends to running work: \texttt{IdeaGo}
preserves completed progress after interruption (OC6), yet the
interrupted run remains ``in progress'' and is not resumed (OC8). The
\texttt{deep-research} agent likewise cannot resume a paused run after
restart or complete a run interrupted by a kill. In contrast,
\texttt{MemCoach} resumes its paused run from state on a bind mount, and
\texttt{LambChat} resumes its interrupted run within 30\,s. Shutdown exposes the same pattern. On both tested systems using
LangGraph's production server, SIGTERM initiates shutdown, but Docker's
10-second stop grace period expires before running work drains (OC4).
Forced recreation interrupts the running worker (OC13), while a
server-process crash leaves the server down without a restart policy
(OC12). For OC5, one system never receives SIGTERM; another receives it
but exceeds the stop grace period, breaking OC4 rather than OC5.

\obs{Two violations predicted from repository evidence do not become
agent failures at runtime because the deployed image provides safeguards
that the repository does not expose.} \texttt{Panaversity} accounts for both cases. Its Compose file neither waits for
database readiness (OC10) nor mounts the state directory used by
\texttt{langgraph dev} (OC2), yet the agent starts after every cold
deployment and preserves every tested conversation across replacement.
The pulled Agent Server image explains both outcomes: it retries the
database and runs migrations before serving, and uses the Postgres-backed
server with conversation state on a named volume. When the development
server runs without this safeguard, as in \texttt{SRE-Agent}, the predicted OC2
failure manifests.

\obs{When an interrupted agent run is resumed, the interrupted step is
re-executed and can therefore repeat its side effects.}
OC4 and OC13 interrupt running work on both tested systems, but the
server subsequently resumes and completes each run 10--221\,s later.
Resumption re-executes the interrupted step: the proxy records its model
call a second time. Thus, recovery can delay completion and can repeat
an external action already performed by that step (OC7).

\obs{Without a mechanism to resume interrupted work, recovery repeats the
entire agent run, while an interrupted stream remains incomplete.}
The \texttt{deep-research} agent exhibits both outcomes. After a kill,
resubmitting its request starts a new run and repeats every model call
made before the interruption, rather than only the interrupted step
(OC7). During streaming, forced recreation terminates the connection;
the API provides no mechanism to resume the stream or retrieve the
remaining output, so the client misses the rest of the answer (OC11).

\newcommand{\addnote}[1]{\textcolor{blue}{\textbf{[TODO:} #1\textbf{]}}}

\section{RQ4: How Do Developers Repair Contract Violations?}
\label{sec:rq4fix}

\para{Motivation.}
RQ2 identifies contract violations in deployed agent systems, and RQ3
shows that they manifest at runtime. RQ4 turns to repository histories
for independent evidence of whether developers encounter these
violations in practice and how they repair them. We examine which
contracts appear in developer repairs, the recurring repair strategies,
and whether each repair satisfies the full deployment obligation.

\para{Approach.}
\underline{(i)~Retrieval.}
Through the GitHub API, we search public commits, pull requests, and
issues using queries that pair \emph{langgraph} with keywords covering
deployment-platform events (e.g., \emph{restart}, \emph{kill}),
agent-framework requirements (e.g., \emph{checkpoint}, \emph{resume}),
and deployment obligations (e.g., \emph{persist}, \emph{retry}).
After independently reading 200 retrieved artifacts, two authors add,
by agreement, developer terms for the same phenomena (e.g.,
\emph{crash}, \emph{stuck}, \emph{orphan}). The queries, run on
2~September 2026, return 39,261 artifacts. We rank artifacts by the number of distinct keywords matched. From the
highest score downward, one author inspects 100 artifacts per score and
stops at the first score with no relevant failure report. The resulting
cutoff retains 7,268 artifacts from 270 repositories. All queries are
in the replication package.

\underline{(ii)~Screening.}
Keyword retrieval favors recall, so artifacts may not
describe relevant failures. Following two-phase
sampling~\cite{cochran1977sampling}, we use GPT-5.4~mini to screen them for a failure following a
deployment-platform event or a change intended to fix or prevent one.
The model serves only as a screening instrument; the authors validate
every retained candidate from the original artifacts and diffs in
step~(iii). Screening yields 396 candidates from 98 repositories. To
assess screening misses, two authors independently inspect a random
sample of 100 rejected artifacts ($\kappa=1$) and find one
missed repair.

\underline{(iii)~Validation.}
A \emph{repair} is a commit, or the commits of a pull request, that
fixes or explicitly prevents a failure caused by a deployment-platform
event. We reconstruct failure--repair pairs through GitHub
cross-references: from an issue to the change that closes or references
it, and from a commit or pull request to the failure described in its
message, linked issue, or diff. Issues without a fixing change, features,
and refactorings are excluded. Two authors independently validate 130
of the 396 candidates (33\%; $\kappa=.85$); disagreements are resolved
by discussion, and the first author applies the resulting rules to the
remainder. This yields 122 repairs from 63
repositories.

\underline{(iv)~Coding.}
We code each repair along three facets. \emph{Contract} identifies the
operational contract whose violation the repair addresses, or
\emph{none} when no contract applies ($\kappa=.91$). \emph{Fix}
captures what the developer changes. Two authors independently describe
each change in the double-coded sample ($\kappa=.92$), then consolidate
the descriptions into recurring fixes and families. The first author
codes the remaining repairs, after which both authors review all
assignments and refine the taxonomy where necessary. We also record
whether each repair modifies application code, deployment files, or
both. \emph{Completeness} records whether the repair satisfies the full
deployment obligation: \emph{yes} if it prevents the failure,
\emph{partly} if it handles the failure or prevents it only under some
conditions, and \emph{no} otherwise ($\kappa=.69$). All agreement values
are computed on the 130 double-coded candidates.

\para{Results.} Table IV lists the fix patterns. 
\obs{Developer repairs map to 12 of the 14 operational contracts and
recur as 35 distinct fix patterns.}
Of the 122 repairs, 111 (91\%) in 58 repositories address a contract
violation. They comprise 35 fix patterns, 15 of which recur in at least
three repositories (Table~\ref{tab:fixrec}); the complete taxonomy is in
the replication package~\cite{MSR2026_replication}. The remaining 11
repairs address failures common to containerized services, such as dead
connections after database restart. No repair addresses OC12 or OC13.

\obs{Developers repair the agent's interrupted work first, its lost
state second, and last the stream a client loses or the resources a
killed run leaves behind (56\%, 35\%, 9\% of the 111 repairs).}
Of the 111 contract repairs, 62 address running work (OC4--OC8, OC14),
39 saved state (OC1--OC3, OC10), and 10 clients or external resources
(OC9, OC11). Resuming interrupted runs (OC8) is most common, with 23
repairs in 15 repositories, followed by state outliving the process
(OC1; 19 in 18), draining running work on shutdown (OC4; 13 in 10), and
memory limits for concurrent runs (OC14; 12 in 10).

\begin{table*}[!t]
\centering
\caption{Recurring repair patterns observed in at least three
repositories (85 repairs across 50 repositories).}
\label{tab:fixrec}
\scriptsize
\setlength{\tabcolsep}{3pt}
\renewcommand{\arraystretch}{1.06}

\begin{tabularx}{\textwidth}{@{}l
  >{\raggedright\arraybackslash}p{2.25cm}
  >{\raggedright\arraybackslash}X
  r
  l
  c@{}}
\toprule
\textbf{Contract} &
\textbf{Deployment-platform event} &
\textbf{Recurring fix} &
\textbf{$n$ (repos.)} &
\textbf{Layer} &
\textbf{Evidence} \\
\midrule

\rowcolor{gray!15}
\multicolumn{6}{@{}l}{\textbf{SAVED STATE}} \\

OC1 &
\multirow[t]{2}{=}{restart, crash, or OOM kill} &
\textbf{F1.1} Replace the in-memory checkpointer with a durable
checkpointer backed by persistent storage. &
12 (12) & code & \cite{fixF11} \\

& &
\textbf{F1.2} If state is absent after restart, reload it from the
checkpointer. &
4 (4) & code & \cite{fixF12} \\

OC2 &
container replacement &
\textbf{F2} Mount the database data directory on a volume so saved state
outlives the container. &
5 (5) & deployment & \cite{fixF21} \\

OC10 &
start ordering &
\textbf{F10} Wait until the database accepts connections before running
setup. &
3 (3) & deployment & \cite{fixF13} \\

\rowcolor{gray!15}
\multicolumn{6}{@{}l}{\textbf{RUNNING WORK}} \\

OC4 &
\multirow[t]{2}{=}{stop} &
\textbf{F4.1} On SIGTERM, stop accepting work and drain running work
before exit. &
7 (6) & code & \cite{fixF41} \\

& &
\textbf{F4.2} Save state before workers stop and bound cleanup to the
stop grace period. &
4 (4) & code & \cite{fixF42} \\

OC5 &
stop &
\textbf{F5} Run the application as PID~1, using \texttt{exec} or an init
process, so SIGTERM reaches it. &
4 (4) & deployment & \cite{fixF5} \\

OC7 &
kill during a step &
\textbf{F7} Guard external actions with an idempotency key or claim so
re-execution does not repeat them. &
5 (5) & code & \cite{fixF71} \\

OC8 &
\multirow[t]{3}{=}{kill, restart, or replacement} &
\textbf{F8.1} Detect interrupted runs and requeue them or mark them
failed. &
9 (7) & code & \cite{fixF81} \\

& &
\textbf{F8.2} At start-up, detect runs left ``running'' and mark them
failed or for review. &
7 (6) & code & \cite{fixF82} \\

& &
\textbf{F8.3} Make recovery tolerate stale markers and partial records
left by an interrupted run. &
4 (3) & code & \cite{fixF83} \\

OC14 &
\multirow[t]{2}{=}{OOM kill} &
\textbf{F14.1} Reduce peak memory use by fixing leaks, paging large
loads, or bounding them. &
8 (7) & code & \cite{fixF201} \\

& &
\textbf{F14.2} Set the memory limit from the workload's measured peak or
bound the process below it. &
4 (4) & deployment & \cite{fixF202} \\

\rowcolor{gray!15}
\multicolumn{6}{@{}l}{\textbf{CLIENTS AND EXTERNAL RESOURCES}} \\

OC9 &
kill &
\textbf{F9} At start-up, find and release external resources left by
interrupted runs. &
6 (5) & code & \cite{fixF91} \\

OC11 &
replacement during streaming &
\textbf{F11} Give the stream an identity and cursor so clients reconnect
from the last delivered event. &
3 (3) & code & \cite{fixF14} \\

\midrule
\multicolumn{3}{@{}l}{\textbf{Total: 15 recurring fixes}} &
\textbf{85 (50)} & & \\
\bottomrule
\end{tabularx}

\smallskip
\begin{minipage}{\textwidth}
\scriptsize
\emph{Note:} $n$ (repos.) = repairs (repositories); 11 recurring fixes
modify application code and four deployment files. The full 35-fix
taxonomy is in the replication package.
\end{minipage}

\end{table*}

\obs{The failures are born in deployment but repaired in code: 78\% of
the repairs change only code, and the few in deployment files repeat
four fixes.} 95 of the 122 repairs (78\%) change code, 4 (3\%) change code and deployment files, and 22 (18\%) change only deployment files; the remaining one has no diff we could inspect. Of the 22, 16 (73\%) apply one of the four fixes made in deployment files (Table~\ref{tab:fixrec}): mounting the database on a volume, running the application as PID~1, waiting for the database before setup, and sizing memory limit.

\obs{Only two thirds of repairs fully satisfy the deployment obligation;
completeness depends on whether the repair requires deployment-platform
support.}
Of the 122 repairs, 80 (66\%) fully satisfy the deployment obligation
and 42 (34\%) do so only partially. Saved-state repairs are usually
complete (32/39, 82\%), because durable storage can be established
directly. Draining running work requires the deployment platform to
allow sufficient time before termination: only 2 of 13 OC4 repairs
provide that time, while the other 11 initiate draining but retain the
10-second stop grace period. Resumption shows a similar gap: 11 of 23
OC8 repairs are partial, including eight that report interrupted runs
rather than resume them.

\section{Threats to Validity}
\label{sec:threats}

\para{Construct validity.}
The contracts are bounded by the pinned documentation, and RQ2 by
repository evidence; behavior hidden inside deployed images can therefore
escape both, as the two RQ3 mismatches show. We mark a contract
\emph{broken} only on positive evidence and \emph{not decidable}
otherwise, making RQ2 violation rates lower bounds. RQ3 mitigates
measurement error through pre-event checks, validated
deployment-platform events, and three repetitions.

\para{Internal validity.}
Manual labels and repair coding involve judgment. We mitigate this
through independent coding, adjudication, codebook refinement, and
reported agreement. LLMs assist candidate extraction (RQ1) and screening
(RQ4), but authors validate every retained candidate; extracted
quotations are also verified verbatim, and a random sample of 100
rejected RQ4 artifacts estimates screening misses. In RQ3, a pinned
model at temperature~0 and a recording proxy control model variation.

\para{External validity.}
We study LangGraph with Docker, Compose, and Kubernetes in public Python
repositories with at least 10 stars; the reported proportions therefore
characterize this population. Other frameworks and deployment platforms
may yield different contracts. RQ3 selects 12 systems for contract
coverage, so it validates runtime manifestation rather than prevalence.
RQ4 is limited by keyword retrieval and by repairs visible in public
repository histories.

\section{Related Work}
\label{sec:related}

\subsection{LLM-based agents and their failures.}

Several empirical studies examine how developers build and operate agent
systems. Chen et al.~\cite{chen2025llmdev} studied the challenges
facing LLM application developers and reported deployment among the
recurring ones; Asgari et al.~\cite{agentchallenges2025} drew the
challenges of agent systems from Stack~Overflow posts and GitHub
issues; Pan et al.~\cite{pan2026measuring} measured agents in
production; further studies examined developer practices, industrial
adoption, configuration, and
observability~\cite{agentpractices2025,apostolou2026industry,galster2026config,rombaut2026scaffold,dong2024agentops}.
A second line of work studies failures within agent systems. Cemri et al.~\cite{cemri2025mast} derived from
multi-agent execution traces a taxonomy of failure modes in how agents
specify, coordinate, and verify their work, and Islam et
al.~\cite{islam2026agentbugs} studied bugs in LLM agents. Other studies examine error propagation and accountability in
role-specialized multi-agent
pipelines~\cite{barrak2025traceability,beem2026broke}. Comparable
studies preceded them for deep-learning
software~\cite{chen2020deploydl,chen2021deployfaults}. 
Our study addresses a
different failure boundary: the interaction between agent-framework
requirements and deployment-platform behavior. Rather than deriving
failures from execution traces or developer reports, we derive
operational contracts from framework and deployment-platform
documentation and examine whether ordinary deployment-platform
events such as restart, replacement, and shutdown violate the
requirements of otherwise correctly executing agents.

\subsection{Container deployment and operational correctness.}

Prior work has extensively studied container deployment artifacts. Studies have characterized Docker usage~\cite{cito2017docker}, identified Docker-specific refactorings and technical debt~\cite{ksontini2021docker,ksontini2024drminer}, and catalogued practices and misconfigurations in Dockerfiles~\cite{henkel2020binnacle,rosa2024dockerfilesmells}, Compose files~\cite{ibrahim2021compose,composepatterns2024}, and Kubernetes manifests~\cite{soldani2020k8smining,neri2020microsmells,rahman2019sins,rahman2023k8smisconfig}. Most closely, Jebari et al.~\cite{jebari2026mlbox} studied 1,993 Dockerfiles from ML projects, characterizing their roles and build inefficiencies. These studies assess deployment artifacts against practices that apply independently of the application they deploy. Our operational contracts instead capture obligations that arise from the interaction between agent code and its deployment environment.

\section{Conclusion}
\label{sec:conclusion}

Container platforms routinely restart, replace, and terminate containers,
while agent systems carry saved state and running work across these
deployment-platform events. Neither the agent framework nor the
deployment platform states the deployment obligations needed when the
two are combined. We introduced 14 \emph{operational contracts} to make
these obligations explicit and derived fourteen for LangGraph with
Docker, Compose, and Kubernetes.

Across 433 deployed agent systems, the boundary is both widespread and
failure-prone: at least one contract applies to 75.5\% of systems, while
37.4\% of those needing durable state lose it on restart and 41.5\% on
container replacement. Execution confirms that these violations matter:
14 of 16 predicted violations manifest as runtime failures, while all
seven scorable predictions that a contract holds are confirmed.
Developer repairs independently cover twelve contracts through 35 fix
patterns; 78\% modify application code, and only two thirds fully
satisfy the deployment obligation.

\bibliographystyle{IEEEtran}
\bibliography{references}

\end{document}